\documentclass[preprint,english,superscriptaddress,showpacs,aps]{revtex4}
\usepackage{amsmath}
\usepackage{amssymb}
\usepackage{graphicx}

\begin{document}

\title{Chaotic traveling rolls in Rayleigh-B\'enard convection}

\author{Supriyo Paul}

\affiliation{Department of Physics and Meteorology, Indian Institute of Technology, Kharagpur-721 302, India}

\affiliation{Department of Physics, Indian Institute of Technology, Kanpur-208 016, India}

\author{Krishna Kumar}

\affiliation{Department of Physics and Meteorology, Indian Institute of Technology, Kharagpur-721 302, India}

\author{Mahendra K. Verma}

\affiliation{Department of Physics, Indian Institute of Technology, Kanpur-208 016, India}

\author{Daniele Carati}

\affiliation{Physique Statistique et Plasmas, Universit\'e Libre de Bruxelles, B-1050 Bruxelles, Belgium}

\author{Arnab K. De}

\affiliation{Department of Mechanical Engineering, Indian Institute of Technology, Kanpur-208 016, India}

\author{Vinayak Eswaran}

\affiliation{Department of Mechanical Engineering, Indian Institute of Technology, Kanpur-208 016, India}

\begin{abstract}
In this paper we investigate two-dimensional (2D) Rayleigh-B\'enard convection using direct
numerical simulation in  Boussinesq fluids of Prandtl number $P = 6.8$ confined
between thermally conducting plates. We show through the simulation  that in a small range
of reduced Rayleigh number $r$ ($770 < r < 890$) the 2D rolls move chaotically in a
direction normal to the roll axis. The lateral shift of the rolls may lead to global flow
reversal  of the convective motion. The chaotic traveling rolls are observed in simulation with 
{\it free-slip} as well as {\it no-slip} boundary conditions on the velocity field. We 
show that the traveling rolls and the flow reversal are due to an interplay between the real and
imaginary parts of the critical modes.
\end{abstract}
 
 
\maketitle
The Rayleigh-B\'enard convection (RBC)~\cite{rayleigh,chandra,busse:1978} is an extensively
studied system for investigating a range of interesting phenomena like instabilities~\cite{ahlers_2000}, pattern-formation~\cite{cross:1993}, chaos~\cite{swinney_gollub:1978, ahlers:1993}, spatio-temporal chaos, and turbulence~\cite{busse:1981,Manneville:book2,ahlers_2006}. The convective flow is characterized by two non-dimensional numbers:  the Rayleigh number $R$ and the Prandtl number $P$. Two-dimensional (2D) stationary rolls~\cite{chandra} 
appear as primary instability in Boussinesq fluids confined between thermally conducting boundaries when the Rayleigh number $R$ is raised just above a critical value $R_c$. The convective dynamics at higher values of R depends on 
Prandl number. Interesting convective dynamics are observed when the Rayleigh number is increased beyond  $R_c$~\cite{ahlers_2000,willis_deardorff:1970,busse_jfm:1972,cleverbusse_jfm:1987,kft:1996}. Busse~\cite{busse_jfm:1972} found traveling waves  along the axis of the cylindrical rolls as secondary instability in low Prandtl number fluids that makes the convection three-dimensional (3D). Traveling waves  are also found  in rotating RBC in cylindrical geometry~\cite{choi_pre:2004}. Another kind of  traveling waves is known to occur in two-dimensional (2D)  RBC in binary mixtures~\cite{walden_prl:1985,barten_prl:1989}, where the straight rolls move in a direction perpendicular to the roll axis. Such behaviour is not reported in two-dimensional RBC in pure fluids, although many interesting features have been investigated 
(e.g.,~\cite{moore_weiss:1973,curry_jfm:1984,thual_jfm:1992}).

In this paper we present results of two-dimensional numerical simulation of
the RBC in a pure Boussinesq fluid of Prandtl number $P = 6.8$. We have used {\it free-slip} as well as {\it no-slip} boundary conditions on
the velocity field at the horizontal plates. In the horizontal direction, we have assumed
{\it periodic} boundary conditions on all the fields.  We report our results as a function of reduced Rayleigh number $r = R/R_c$. For {\it free-slip} boundaries,  at $r \approx 130$ we observe a transition from one oscillatory state with only real 
or imaginary values of critical modes $W_{101}$ and $\theta_{101}$ to another oscillatory 
state where both the real and imaginary parts are non-zero.
At later $r$, near $770 < r < 890$ we find chaotic traveling waves in a direction normal to the roll-axis.  
The sudden lateral shift in the roll system leads to either jitters in the convective flow or  flow 
reversal of the convective motion. The chaotic travelling waves are also observed in the direct 
numerical simulation (DNS) of the RBC with {\it no-slip} boundaries. We find links between the chaotic 
traveling waves or rolls and the flow reversal.

We consider 2D convection in an extended  layer of Boussinesq fluid with thermal expansion coefficient $\alpha$, kinematic viscosity $\nu$, thermal diffusivity $\kappa$ that is enclosed between two flat conducting plates separated by 
distance $d$ and heated from below. The adverse temperature gradient  is $\Delta T/d$, where 
$\Delta T$ is the temperature difference imposed across the bounding plates.  The nondimensional equations are
\begin{eqnarray*}
\frac{\partial\mathbf{u}}{\partial t}+(\mathbf{u}\cdot\nabla)\mathbf{u} & = & -\nabla \sigma + RP\theta\hat{{\mathbf{z}}}+P\nabla^{2}\mathbf{u},\\
\frac{\partial\theta}{\partial t}+(\mathbf{u}\cdot\nabla)\theta & = & u_{3}+\nabla^{2}\theta. \end{eqnarray*}
Here the Rayleigh number 
$R = \frac{\alpha g (\Delta T) d^3}{\nu \kappa}$ is the ratio of the buoyancy and the dissipative forces, while 
the Prandtl number $P = \nu/\kappa$ is the ratio of thermal diffusive time $\tau_{th} = d^2/\kappa$ and the 
viscous diffusive time $\tau_{vis} = d^2/\nu$. The temperature perturbation $\theta$ due to the convective flow vanishes
at thermally conducting boundaries. The realistic {\it no-slip} boundary conditions imply  that the velocity field 
${\bf v} = (u, 0, w) = 0$ at the boundaries. The idealized {\it free-slip} boundaries imply 
$\partial_z u  = w =0$ at the flat plates. The value of $R_c$ is $ 27 \pi^4/4 \approx 657.5$ 
for {\it free-slip} boundaries, and  $R_c \approx 1707.8$ for {\it no-slip} boundaries.

We have carried out  
our simulations of the RBC in 2D using pseudo-spectral method~\cite{thual_jfm:1992}
with stress-free boundaries. The grid resolution used for the simulations with the 
{\it free-slip} boundaries is $256\times 256$  with an
aspect ratio of $2\sqrt{2} : 1$. The reduced Rayleigh number $r=R/R_c$ is varied from $1.01$
to $10^3$.  We have employed fourth-order Runge-Kutta (RK4) scheme for time stepping with the
time step varying from $1\times10^{-4}$ to $1\times10^{-6}$ in thermal diffusive time units. 
The code was validated using the results reported by Thual~\cite{thual_jfm:1992}. 

\begin{figure}[t]
\includegraphics[height=!,width=3.5in]{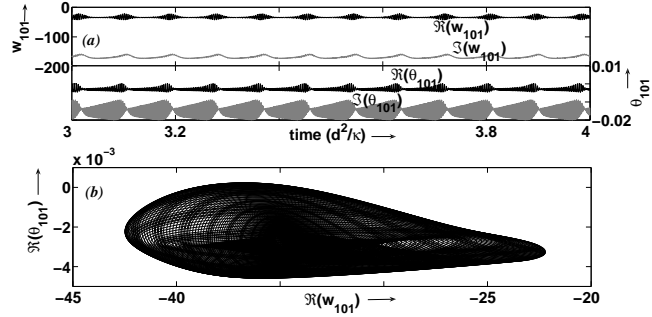}
\caption{(a) Timeseries of $\Re(W_{101})$, $\Im(W_{101})$, $\Re(\theta_{101})$, $\Im(\theta_{101})$ at reduced Rayleigh number $r=700$ and Prandtl number $P=6.8$.  The timeseries has two leading frequencies, and it shows quasiperiodic behaviour. (b) The projection of the phase space in ${\Re(W_{101})} - {\Re(\theta_{101})}$ that elucidates the quasiperiodic behaviour of the system.}
\label{fig:QP}
\end{figure}

To solve {\em no-slip} convective flow, we employed finite-difference two-step precedure with the Adam-Bashforth-Crank-Nicolson scheme.  The first step involves pressure calculation using Poisson's equation, and the second step deals with time-advancing using fourth-order central explicit scheme \cite{Tam} with enhanced spectral resolution.  We computed the no-slip flow for $r=830$ and $P=6.8$ and compared the results with that obtained with free-slip simulation.

First we report the results of 2D simulation with free-slip boundary conditions.
We observe long time behaviour of the critical 2D modes $W_{101}$ and $\theta_{101}$. 
Here the three subscript denote the wavenumber indices of the Fourier modes along $x$, $y$,  and $z$ directions, where $z$ is the vertical direction.  In our simulation the second subscript is zero.  We start our simulation with real Fourier modes.  For reduced Rayleigh number $r \le 80$, the system shows time independent steady convection.  For $80 < r < 130$, we observe simple oscillations in all the modes. Here all the Fourier modes remain real. Similar observations have been made previously by others~\cite{thual_jfm:1992,moore_weiss:1973,curry_jfm:1984}. As $r$ is raised further, the critical Fourier modes ($W_{101}$ and $\theta_{101}$) become complex. For $130 < r< 145$, the imaginary  parts of $W_{101}$ and $\theta_{101}$ become nonzero and they oscillate around a zero mean,
while the real parts oscillate around a nonzero mean. 

For $145 < r < 770$ the  imaginary parts of the complex modes also oscillate around a finite mean. This is an oscillatory instability, which makes the critical modes complex. The real and imaginary parts of a Fourier mode interact with 
each other through nonlinear interaction with higher order modes.  When the simulation is
started with purely imaginary modes (except $\theta_{00n}$, which is always real due to the reality condition), then the roles of real and imaginary parts of all complex modes get interchanged. 

\begin{figure}
\includegraphics[height=!,width=3.5in]{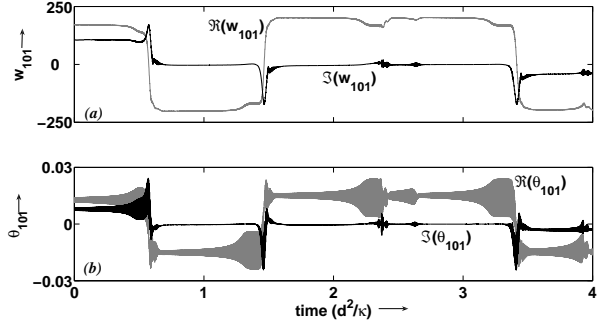}
\caption{Timeseries of the real and imaginary parts of the critical velocity mode  $W_{101}$ (a), and the  temperature mode $\theta_{101}$ (b) for $r = 830$. The origin on dimensionless time axis is suitably chosen after the steady-state has been reached.  The timeseries of the modes shows jitters (small fluctuations) and flips.  The fluctuations in the temperature mode is stronger than in that in the velocity mode.}
\label{fig:chaos_timeseries}
\end{figure}

\begin{figure}[t]
\includegraphics[height=!,width=3.5in]{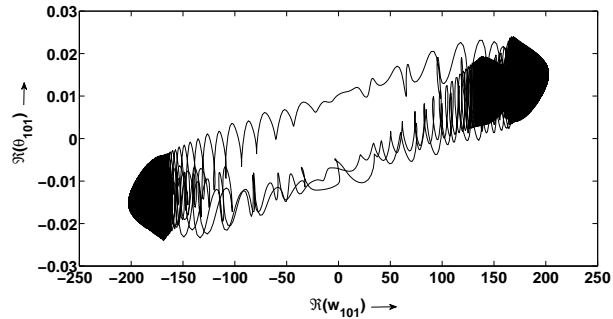}
\caption{Projection of the phase space in the ${\Re(W_{101})} - {\Re(\theta_{101})}$ plane
at $r = 830$. The figure illustrates the chaotic behaviour of the system.}
\label{fig:chaos_phasespace}
\end{figure}

The oscillations of both real and imaginary parts become strongly anharmonic for
$660 < r < 770$.  This leads to temporally quasiperiodic flow. Fig.~\ref{fig:QP}(a) 
illustrates the real and the imaginary parts of the critical mode $W_{101}$ and $\theta_{101}$. These modes
show two leading frequencies. The projection of phase space on $W_{101} - \theta_{101}$ 
shows quasiperiodic flow as seen in Fig.~\ref{fig:QP}(b). Similar behavior has been observed  
in experiments on RBC by Swinney and Gollub~\cite{swinney_gollub:1978}.

\begin{figure*}[t]
\includegraphics[height=!,width=6.5in]{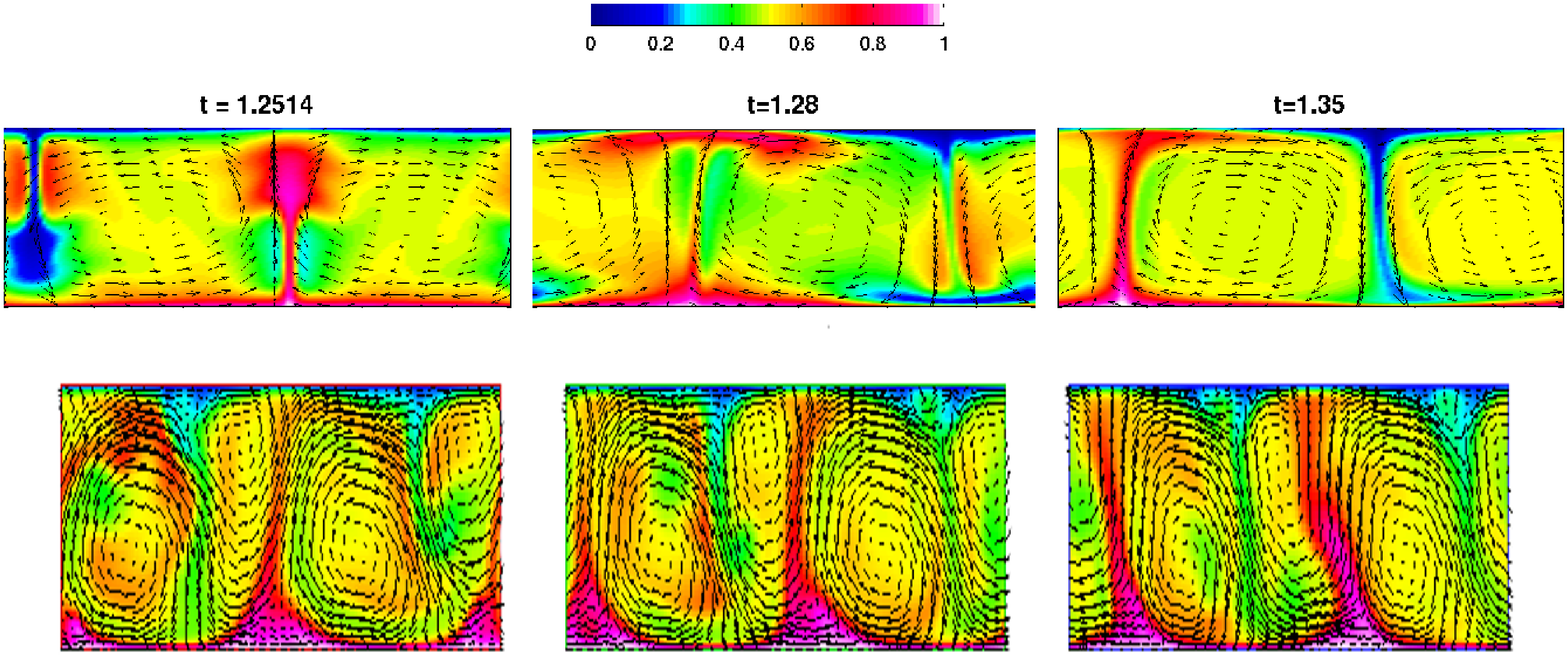}
\caption{Travelling rolls in two-dimensional Rayleigh-B\'enard convection at $r = 830$.  
The top panel of the figures is for the simulation with {\it free-slip} boundaries and the 
bottom panel is for the {\it no-slip} boundaries. The temperature field is represented in color with blue being the coldest region.  The arrows represent the velocity field.  The three images at the top panel show the field configurations at three time. At $t=1.2514$ the central region has hot fluid that shifts leftward subsequently.  By $t=1.35$ the whole roll pattern is shifted approximately by half a wavelength.   Similar features are seen for no-slip convective flow, except that the flow pattern moves right.}

\label{fig:flow_reversal}
\end{figure*}

The convection becomes  chaotic for $770 < r < 890$. Figure~\ref{fig:chaos_timeseries} displays 
the real and the imaginary parts of the critical modes $W_{101}$ and $\theta_{101}$ for
$r = 830$. The projection of the phase space on ${\Re(W_{101})} - {\Re(\theta_{101})}$
plane is illustrated in Fig.~\ref{fig:chaos_phasespace} that indicates the chaotic nature of the system.

In the chaotic regime, the real and imaginary parts of the critical modes interact strongly through nonlinear coupling
with higher order Fourier modes.  As illustrated in Fig.~\ref{fig:chaos_timeseries},  the Fourier modes show jitters (small fluctuations) as well as change in sign.   The velocity mode $W_{101}$  and temperature mode $\theta_{101}$  appear to be in approximate phase with each other.  However the fluctuations in the temperature mode near the jitter or the flip is stronger than the corresponding fluctuations in the velocity mode. The interval between two flips is of the order of one thermal diffusion time scale, while the duration of the flips is much shorter.

The physical interpretation of the chaotic timeseries (Fig.~\ref{fig:chaos_timeseries}) yields interesting insights into the traveling wave instability and flow reversal.   We provide a simple argument to explain this. If the critical modes were the only modes in the system, then the vertical velocity in real space would be $\Re(W_{101} \exp(i k_c x) \sin(n \pi z))$.  When $W_{101}$ changes from a real value $A$ to a complex value $A \exp(i \phi)$, it would lead to a shift of the roll pattern in horizontal distance by $\phi/k_c$.  This is the observed {\em traveling wave instability} near the bifurcation. 
Since the temporal change in the phase of the critical mode is chaotic, the movement of the rolls would also be irregular. 
During the flip of the critical Fourier modes, $\phi \approx \pi$, which corresponds to a horizontal shift of $\lambda/2$. Under this situation, the vertical velocity reverses its direction at a given location that leads to global flow reversal in the system.  The simulated flow contains many modes apart from the critical modes, yet the large-scale ones play a dominant role.  Hence, the bifurcation of the large-scale Fourier modes from real to complex play a major role in initiating traveling wave instability and flow reversal.

We illustrate the above dynamics using Fig.~\ref{fig:flow_reversal} that  contains two panels of images: the top ones are for the {\it free-slip} boundaries, while the bottom ones are for the {\it no-slip} boundaries.    The two panels of Fig.~\ref{fig:flow_reversal} illustrate three snap-shots of the velocity and temperature at three different instances of time.  The top panel (free-slip) illustrates the flow fields at $t=1.2514$,  $t=1.28$, and $t=1.35$ respectively for a different run.  The temperature fields are shown in color coding with blue being the coldest region, while the velocity fields are shown using arrows.    At $t=1.2514$, the hot fluid is rising near the center.  The roll pattern moves leftward as time advances.  At time $t=1.35$, a cold fluid parcel is falling down near the center.  Thus the top panel illustrates the traveling roll in RB convection under free-slip boundary condition.   In the bottom panel, similar feature is observed for no-slip boundary condition however for a shorter time.  Here the rolls have moved only a smaller fraction of a wavelength.  Note that the rolls are traveling  perpendicular to the roll axis along the periodic direction in both the results, and the interval between two consecutive flow reversal is of the order of one diffusive time unit (refer to Fig.~\ref{fig:chaos_timeseries}).

Figure \ref{fig:flow_reversal} also depicts flow reversal phenomenon.   If we place a probe near the center of the system, we would observe the fluid to be hot and moving upward at initial time.   At the end, the above mentioned probe would measure colder fluid flowing downward.  Thus a flow reversal due to traveling rolls is clearly illustrated.  

We have created two movies to illustrate the traveling rolls and flow reversal by taking many frames between the reversal.  These movies can be seen in our website~\cite{movie:free_slip, movie:no_slip} . The first movie~\cite{movie:free_slip} depicts the traveling rolls near a flow reversal regime.    By the end of the movie (in approximately 0.1  thermal diffusive time unit), the roll has shifted approximately by half a wavelength.     Similar features are observed in the second movie~\cite{movie:no_slip} that depicts the convective flow with no-slip boundaries; here the roll pattern moves to the right.  

We have performed several three-dimensional (3D) simulations for Prandtl number 6.8, and a range of Rayleigh number in the chaotic and turbulent regime.  We observe the above features of traveling rolls and flow reversal in 3D as well; these reults will be reported in due course.

Reversal of large-scale flows has been observed in many experiments and numerical simulations \cite{KrisHowa:PNAS,KRS:nature_wind,Cioni,Qui,Ahle:JFM}.  Some of the experiments have been performed for large Rayleigh numbers (of the order of $10^9$ to $10^{12}$).   The physics of the above phenomenon has been of great interest in the recent past.  Our results are for two dimensional RBC  at moderately low Rayleigh numbers.  Yet, the simple mechanism of traveling rolls in the periodic direction and its consequence to the phenomenon of flow reversal may be of interest in more complex situations.   The large Rayleigh number regimes are under investigation.

To summarize, we observe succession of patterns in two-dimensional RBC under free-slip boundaries.  Our focus has been on the chaotic traveling rolls observed in the range of $770 < r < 890$.  We show that the generation of traveling rolls is due to the bifurcation of critical modes from real values to complex values.   This kind of traveling roll is possible along the periodic direction that may be realized in practice along the azimuthal direction in a cylindrical container.  It is interesting to note that several traveling roll states and flow reversals have been observed in cylindrical geometry.   Also, the chaotic traveling rolls lead to flow reversal when the amplitude of the critical Fourier modes switches sign.

\begin{acknowledgments}
We thank Stephan Fauve, Pankaj Mishra, K. R. Sreenivasan, J. Niemela, and Bruno Eckhardt for discussions and  various important suggestions.  Part of the work was supported by the grant of Swarnajayanti fellowship by Department of Science and Technology, India.
\end{acknowledgments}

\bibliographystyle{apsrev}
%\bibliography{/home/mkv/res/bib/res}

\end{document}